\documentclass[reqno,centertags, 12pt]{amsart}
\usepackage{amsmath,amsthm,amscd,amssymb}
\usepackage{latexsym, verbatim}

\newcommand{\bbR}{{\mathbb{R}}}

\newcommand{\bbZ}{{\mathbb{Z}}}

\newcommand{\lb}{\label}
\newcommand{\f}{\frac}

\newcommand{\bi}{\bibitem}

\newcommand{\beq}{\begin{equation}}
\newcommand{\eeq}{\end{equation}}
\newcommand{\ba}{\begin{align}}
\newcommand{\ea}{\end{align}}
\newcommand{\eps}{\varepsilon}
\newcommand{\tht}{\theta}

\newcounter{smalllist}
\newenvironment{SL}{\begin{list}{{\rm\roman{smalllist})}}{%
\setlength{\topsep}{0mm}\setlength{\parsep}{0mm}\setlength{\itemsep}{0mm}%
\setlength{\labelwidth}{2em}\setlength{\leftmargin}{2em}\usecounter{smalllist}%
}}{\end{list}}

 \DeclareMathOperator{\Tr}{Tr}

 \allowdisplaybreaks

\newtheorem{theorem}{Theorem}

\newtheorem{lemma}[theorem]{Lemma}
\newtheorem{corollary}[theorem]{Corollary}
\theoremstyle{definition}

\theoremstyle{remark}

\begin{document}
\title[Sum Rules and Divergent Lieb-Thirring Sums]{Sum Rules for Jacobi Matrices and Divergent Lieb-Thirring Sums}
\author[Andrej Zlato\v{s}]{Andrej Zlato\v{s}}

\address{ Department of Mathematics \\ University of
Wisconsin \\ Madison, WI 53706, USA \\ {\it E-mail address}: \tt
andrej@math.wisc.edu}




\begin{abstract} Let $E_j$ be the
eigenvalues outside $[-2,2]$ of a Jacobi matrix with
$a_n-1\in\ell^2$ and $b_n\to 0$, and $\mu'$ the density of the
a.c.~part of the spectral measure for the vector $\delta_1$. We
show that if $b_n\notin\ell^4$, $b_{n+1}-b_{n}\in\ell^2$, then
\[
\sum_j(|E_j|-2)^{5/2} = \infty,
\]
and if $b_n\in\ell^4$, $b_{n+1}-b_{n}\notin\ell^2$, then
\[
\int_{-2}^2 \ln(\mu'(x)) (4-x^2)^{3/2}\, dx = -\infty.
\]
We also show that if $a_n-1,b_n\in\ell^3$, then the above integral
is finite if and only if $a_{n+1}-a_{n}, b_{n+1}-b_{n}\in\ell^2$.
We prove these and other results by deriving sum rules in which
the a.c.~part of the spectral measure and the eigenvalues appear
on opposite sides of the equation.
\end{abstract}

\maketitle

\section{Introduction} \lb{s1}

In the present paper we consider Jacobi matrices
\begin{equation*}
J\equiv \begin{pmatrix}
b_1 & a_1 & 0 & \dots \\
a_1 & b_2 & a_2 & \dots \\
0 & a_2 & b_3 & \dots \\
\dots & \dots & \dots & \dots \\
\end{pmatrix}
\end{equation*}
with $a_n>0$, $b_n\in\bbR$, and $a_n\to 1$, $b_n\to 0$. These are
compact perturbations of the {\it free matrix} $J_0$ with
$a_n\equiv 1$ and $b_n\equiv 0$. If only $a_n\equiv 1$, then $J$
is the discrete half-line Schr\" odinger operator with the
decaying potential $b_n$.

$J$ is a self-adjoint operator acting on $\ell^2(\{1,2,\dots\})$.
We denote by $\mu$ the spectral measure of the (cyclic for $J$)
vector $\delta_1$ and by $\mu'$ the density of its a.c.~part. For
$J_0$, the measure $\mu_0$ is absolutely continuous with $\mu_0'
(x)=(2\pi)^{-1} \sqrt{4-x^2} \chi_{[-2,2]}(x)$, and so by Weyl's
theorem, $\sigma_{\rm ess}(J)=\sigma_{\rm ess}(J_0)=[-2,2]$.
Hence, outside $[-2,2]$ spectrum of $J$ consists only of
eigenvalues (of multiplicity 1), with $\pm 2$ the only possible
accumulation points. We will denote the negative ones
$E_1,E_3,\dots$ and the positive ones $E_2,E_4,\dots$, with the
convention that $E_{2j-1}\equiv -2$ ($E_{2j}\equiv 2$) if $J$ has
fewer than $j$ eigenvalues below $-2$ (above $2$).

We let $\partial a_n\equiv a_{n+1}-a_{n}$, $\partial b_n\equiv
b_{n+1}-b_{n}$, and define
\[
r_n\equiv b_n^4-2(\partial b_n)^2 -8(\partial a_n)^2+
4(a_n^2-1)(b_n^2+b_nb_{n+1}+b_{n+1}^2).
\]
The following are our main results.

\begin{theorem} \lb{T.1}
Assume $a_n-1\in\ell^3$ and $b_n\to 0$.
\begin{SL}
\item[{\rm{(i)}}] If $\sum_{n=1}^\infty r_n=\infty$ or does not
exist, then $\sum_{j=1}^\infty (|E_j|-2)^{5/2} = \infty$.
\item[{\rm{(ii)}}] If $\sum_{n=1}^\infty r_n=-\infty$ or does not
exist, then $\int_{-2}^2 \ln(\mu'(x)) (4-x^2)^{3/2}\, dx =
-\infty$.
\end{SL}
\end{theorem}

{\it Remark.} One can actually dispense with the assumption
$a_n-1\in\ell^3$, but the corresponding $r_n$ is less transparent
(it is the diagonal element of the matrix $P_w(J)$ from the proof
of Theorem \ref{T.1}).

\begin{corollary} \lb{C.1a}
Assume $a_n-1\in\ell^2$ and $b_n\to 0$.
\begin{SL}
\item[{\rm{(i)}}] If $b_n\notin\ell^4$ and $\partial
b_n\in\ell^2$, then $\sum_{j=1}^\infty (|E_j|-2)^{5/2} = \infty$.
\item[{\rm{(ii)}}] If $b_n\in\ell^4$ and $\partial
b_n\notin\ell^2$, then $\int_{-2}^2 \ln(\mu'(x)) (4-x^2)^{3/2}\,
dx = -\infty$.
\end{SL}
\end{corollary}

\begin{proof}
Since $a_n-1\in\ell^2$, we have $\partial a_n\in\ell^2$. Also,
\[
|4(a_n^2-1)(b_n^2+b_nb_{n+1}+b_{n+1}^2)|\le 72 (a_n^2-1)^2 +
\tfrac 14(b_n^4+b_{n+1}^4)
\]
and $a_n^2-1\in\ell^2$, so the result follows from Theorem
\ref{T.1}.
\end{proof}

The sum in (i) is a Lieb-Thirring sum and most results go in the
direction opposite to (i), bounding sums of eigenvalue moments
from above in terms of the matrix elements. See \cite{HS}, where
it is proved that
\begin{equation} \lb{1.0a}
\sum_j(|E_j|-2)^{p} \le c_p \Big( \sum_n|a_n-1|^{p+1/2} +
\sum_n|b_n|^{p+1/2} \Big)
\end{equation}
for any $p\ge \tfrac 12$ and some $c_p>0$, and references therein.

The integral in (ii) is one from a family of Szeg\H o-type
integrals recently studied, among others, in
\cite{DHS,KS,Kupin2,Kupin,LNS,NPVY,SZ,Zla}. The actual Szeg\H o
integral has the weight $(4-x^2)^{-1/2}$ instead of
$(4-x^2)^{3/2}$ and is an important object in the theory of
orthogonal polynomials.

We also single out the following of our results (cf. Corollary
\ref{C.7}).

\begin{theorem} \lb{T.1b}
Assume $a_n-1,b_n\in\ell^3$. Then $\partial a_n,\partial
b_n\in\ell^2$ if and only if $\int_{-2}^2 \ln(\mu'(x))
(4-x^2)^{3/2}\, dx >-\infty$.
\end{theorem}

{\it Remarks.} 1. The ``only if'' part was proved in
\cite{Kupin2}.
\smallskip

2. Note that $a_n-1,b_n\in\ell^3$ and \eqref{1.0a} imply
$\sum_{j=1}^\infty (|E_j|-2)^{5/2} < \infty$.
\smallskip


We briefly review here related results. In \cite{KS}, which
started recent development in the area of sum rules, it is proved
that $a_n-1,b_n\in\ell^2$ if and only if $\sum_j(|E_j|-2)^{3/2} <
\infty$ and $\int_{-2}^2 \ln(\mu'(x)) (4-x^2)^{1/2}\, dx >
-\infty$. Using a higher sum rule \cite{Kupin} shows that
$a_n-1,b_n\in\ell^4$ and $\partial^2a_n,\partial^2b_n\in\ell^2$ if
and only if $\sum_j(|E_j|-2)^{7/2} < \infty$ and $\int_{-2}^2
\ln(\mu'(x)) (4-x^2)^{5/2}\, dx > -\infty$. Finally, \cite{LNS}
shows that $a_n-1,b_n\in\ell^4$ and
$a_{n+1}+a_n,b_{n+1}+b_n\in\ell^2$ if and only if
$\sum_j(|E_j|-2)^{3/2} < \infty$ and $\int_{-2}^2 \ln(\mu'(x))
x^2(4-x^2)^{1/2}\, dx > -\infty$. Closely related to our work is
also a general ``existence'' result in \cite{NPVY}.

From most such results one can conclude that a Lieb-Thirring sum
or a Szeg\H o-type integral is infinite for certain $a_n,b_n$, but
is not able to say which one of these happens. We achieve this by
obtaining sum rules in which these two quantities appear on
opposite sides of the equation (Theorem \ref{T.5}(i)). This is in
the spirit of Theorem 4.1 in \cite{SZ}, which shows that
$\limsup_n\sum_{j=1}^n\ln(a_n)=\infty$ implies
$\sum_j(|E_j|-2)^{1/2} = \infty$ and
$\liminf_n\sum_{j=1}^n\ln(a_n)=-\infty$ implies $\int_{-2}^2
\ln(\mu'(x))(4-x^2)^{-1/2}\, dx = -\infty$ (see also Theorem
\ref{T.8} below).



The paper is organized as follows. In Section \ref{s2} we
introduce the necessary tools, {\it Case sum rules for Jacobi
matrices} (see \cite{Case,KS,SZ,Zla}), and then extend these to a
form we will need here (Theorem \ref{T.5}). In Section \ref{s3} we
use them to prove Theorems \ref{T.1} and \ref{T.1b} and related
results. The author would like to thank Barry Simon for useful
communication.

\section{Sum Rules for Jacobi Matrices} \lb{s2}

In this section we use the notation of and extend results from
\cite{SZ}. If for some $\{c_\ell\}_{\ell=0}^k$ we have
$w(\tht)\equiv \sum_{\ell=0}^k c_\ell\cos(\ell\tht)\ge 0$, we
define
\begin{align}
Z_w (J) &\equiv -\f{1}{2\pi} \, \int_0^{\pi} \ln \biggl(
\f{\pi\mu'(2\cos\tht)}{\sin\tht}\biggr)\, w(\tht) \, d\theta
\notag
\\ &= -\f{1}{2\pi} \int_{-2}^2 \ln \biggl( \f{\mu'(x)}{\mu_0'(x)}\biggr)
 \sum_{\ell=0}^k c_\ell T_\ell \Big(\frac x2 \Big)
\f{dx}{\sqrt{4-x^2}}, \lb{2.1}
\end{align}
where $T_\ell(\cos\tht)\equiv \cos\ell\tht$ is the $\ell^{\rm th}$
Chebyshev polynomial (of degree $\ell$), and the second equality
follows from the substitution $x=2\cos\tht$. Since
$\ln(\mu'(x))\le \mu'(x)\sqrt{4-x^2}-\ln(\sqrt{4-x^2})$,
$\mu(\bbR)=1$, and $w(\tht)\ge 0$, the positive part of the
integral is bounded, that is,
\begin{equation} \lb{2.2}
Z_w(J)\ge C_w
\end{equation}
with $C_w>-\infty$ (but $Z_w(J)=\infty$ is possible). Note that
$Z_w(J_0)=0$.

We also let $|\beta_j|\ge 1$ be such that
$E_j=\beta_j+\beta_j^{-1}$. Hence
\begin{equation*} 
|\beta_j|-1=(|E_j|-2)^{1/2} + O(|E_j|-2).
\end{equation*}

We define
\begin{equation*}
J^{(n)}= \begin{pmatrix}
b_{n+1} & a_{n+1} & 0 & \dots \\
a_{n+1} & b_{n+2} & a_{n+2} & \dots \\
0 & a_{n+2} & b_{n+3} & \dots \\
\dots & \dots & \dots & \dots \\
\end{pmatrix}
\end{equation*}
that is, $J^{(n)}$ is the matrix we obtain from $J$ by removing
the first $n$ rows and columns. We let $E_j^{(n)}\equiv
E_j(J^{(n)})$ and $\beta_j^{(n)}\equiv \beta_j(J^{(n)})$. We also
let $J_n$ be the matrix one obtains from $J$ by replacing $a_j$ by
1 and $b_{j+1}$ by 0 for $j\ge n$. Notice that $(J_n)^{(n)}=J_0$.

We denote (with $\ell\ge 1$)
\begin{align*}
X_0^{(n)}(J) &\equiv \sum_{j=1}^\infty  \Big[
\ln(|\beta_j|)-\ln(|\beta_j^{(n)}|) \Big], 
\\ X_\ell^{(n)}(J) &\equiv \frac 1{2\ell} \sum_{j=1}^\infty  \Big[
\big( \beta_j^\ell-\beta_j^{-\ell} \big)- \big(
(\beta_j^{(n)})^\ell-(\beta_j^{(n)})^{-\ell} \big)  \Big].  
\end{align*}
These sums are always convergent because
$X_\ell^{(n)}(J)=\sum_{j=0}^{n-1} X_\ell^{(1)}(J^{(j)})$ (where
$J^{(0)}\equiv J$), and the finiteness of $X_\ell^{(1)}(J)$
follows from the fact that positive (resp. negative) eigenvalues
of $J$ and $J^{(1)}$ interlace \cite{SZ}.

Finally, if $B$ is a semi-infinite matrix, we let $B(n)$ be the
matrix we obtain from $B$ by adding to it, from the top and left,
$n$ rows and columns containing only zeros. For instance,
$J^{(n)}(n)$ is the matrix one obtains from $J$ by replacing
$a_j,b_j$ for $j\le n$ by zeros. We then define
\begin{align*}
\xi_0^{(n)}(J) &\equiv -\sum_{j=1}^n \ln(a_j) 
\\ \xi_\ell^{(n)}(J) &\equiv -\frac 1{\ell} \Tr \Big( T_\ell \big(\tfrac 12 J \big) - T_\ell \big(\tfrac 12 J^{(n)} \big)(n) \Big) 
\end{align*}
for $\ell\ge 1$. These are well defined because the diagonal of
the matrix $T_\ell(\tfrac 12 J) - T_\ell(\tfrac 12 J^{(n)})(n)$
eventually vanishes (starting from $(n+\ell)^{\rm th}$ diagonal
element), although the matrix need not be trace class.

With this notation one has the following {\it step-by-step sum
rule}:

\begin{lemma} \lb{L.2}
If $w(\tht)=\sum_{\ell=0}^k c_\ell\cos(\ell\tht)\ge 0$, then
\begin{equation} \lb{2.3}
Z_w(J)=\sum_{\ell=0}^k c_\ell\xi_\ell^{(n)}(J) + \sum_{\ell=0}^k
c_\ell X_\ell^{(n)}(J) + Z_w(J^{(n)}).
\end{equation}
\end{lemma}

Remark. Here both sides can be $+\infty$. In particular,
$Z_w(J)=\infty$ if and only if $Z_w(J^{(n)})=\infty$.

\begin{proof}
One proves the statement for $n=1$ and then iterates the obtained
formula $n$ times. The proof is identical to that of Theorems
3.1--3.3 in \cite{SZ} (using their Remark 1 before Theorem 2.1),
where $w(\tht)\equiv 1$, $w(\tht)\equiv 1\pm\cos\tht$, and
$w(\tht)\equiv 1-\cos 2\tht$ (the proofs, with more detail, also
appear in \cite{Zla}).
\end{proof}

A natural question here is what happens when we take $n\to\infty$.
To do this we need to determine the convergence of the terms on
the right hand side of \eqref{2.3}. Following \cite{SZ}, one can
use two approximations --- $J$ by $J_n$, and $J_0$ by $J^{(n)}$
--- to treat the second and third term. We define
\begin{equation} \lb{2.4}
 f_w(\beta)\equiv c_0\ln(|\beta|)+\sum_{\ell=1}^k
\frac{c_\ell}{2\ell} \big( \beta^\ell-\beta^{-\ell} \big)
\end{equation}
so that
\[
\sum_{\ell=0}^k c_\ell X_\ell^{(n)}(J) = \sum_{j=1}^\infty \Big[
f_w(\beta_j)-f_w(\beta_j^{(n)}) \Big].
\]
We also let $X_{\ell,+}^{(n)}(J)$ be defined as $X_\ell^{(n)}(J)$
but with the sum taken only over positive eigenvalues. Similarly
we define $X_{\ell,-}^{(n)}(J)$, with only negative eigenvalues.
We then have
$X_\ell^{(n)}(J)=X_{\ell,+}^{(n)}(J)+X_{\ell,-}^{(n)}(J)$ and
\begin{align*}
\sum_{\ell=0}^k c_\ell X_{\ell,+}^{(n)}(J) & = \sum_{E_j\ge 2}
\Big[ f_w(\beta_j)-f_w(\beta_j^{(n)}) \Big],
\\ \sum_{\ell=0}^k c_\ell X_{\ell,-}^{(n)}(J) & = \sum_{E_j\le -2} \Big[
f_w(\beta_j)-f_w(\beta_j^{(n)}) \Big].
\end{align*}

\begin{lemma} \lb{L.3}
If $a_n\to 1$, $b_n\to 0$, and $w(\tht)=\sum_{\ell=0}^k
c_\ell\cos(\ell\tht)\ge 0$, then
\begin{align}
\liminf_{n\to\infty} Z_w(J^{(n)}) & \ge Z_w(J_0)=0, \lb{2.5}
\\ \liminf_{n\to\infty} Z_w(J_n) & \ge Z_w(J). \lb{2.6}
\end{align}
Also,
\begin{align}
\lim_{n\to\infty}\sum_{\ell=0}^k c_\ell X_{\ell,\pm}^{(n)}(J)  &
=\sum_{\pm E_j\ge 2} f_w(\beta_j),  \lb{2.7}
\\ \lim_{n\to\infty}\sum_{\ell=0}^k c_\ell X_{\ell,\pm}^{(n)}(J_n)
 & =\sum_{\pm E_j\ge 2} f_w(\beta_j). \lb{2.8}
\end{align}
\end{lemma}

{\it Remarks.} 1. Eqs. \eqref{2.7}, \eqref{2.8} are intended as
two statements each --- one with the plus signs and one with the
minus signs. This will be the case in Theorems \ref{T.5}(ii) and
\ref{T.8} as well.

2. The sums on the left hand sides of \eqref{2.7}, \eqref{2.8}
both exist but could be $\pm\infty$. We separate the sums over
positive and negative eigenvalues from each other because one
could be $\infty$ and the other $-\infty$.

\begin{proof}
Eqs. \eqref{2.4}, \eqref{2.5} follow directly from Corollary 5.3
in \cite{KS}.

Let us prove \eqref{2.7}, \eqref{2.8} with the plus signs (the
second case is identical). Notice that $f_w$ is continuous on
$[1,\infty)$ with $f_w(1)=0$. Since also $f_w\in C^\infty$ and not
all its derivatives at $1$ vanish (unless $f_w\equiv 0$), it is
monotone on some interval $[1,1+\eps]$, $\eps>0$ (and so the sums
in \eqref{2.7}, \eqref{2.8} exist). For such functions \eqref{2.7}
holds by Lemma 4.6 in \cite{SZ}. Similarly, \eqref{2.8} holds by
Theorem 6.2 in \cite{KS}, using that $(J_n)^{(n)}=J_0$ has no
eigenvalues (and so the left hand side is just $\lim_{n\to\infty}
\sum_{E_j\ge 2} f_w(\beta_j(J_n))$).
\end{proof}

To treat the first sum in \eqref{2.3} we define
\begin{equation} \lb{2.9}
P_w(J)\equiv S-c_0 A-\sum_{\ell=1}^k \frac {c_\ell}\ell
T_\ell(\tfrac 12 J)
\end{equation}
where $A$ is the matrix with $\ln(a_j)$ on the diagonal, and $S$
is the matrix with $S_{1,1}=-\sum_{\ell=1}^k
\tfrac{1}{4\ell}(1+(-1)^\ell)c_\ell$ and all other elements zero.

\begin{lemma} \lb{L.4} If $n>k$, then with $o(1)=o(n^0)$,
\begin{equation} \lb{2.10}
\sum_{\ell=0}^k c_\ell\xi_\ell^{(n)}(J) = \sum_{j=1}^{n} \big(
P_w(J) \big)_{j,j} +  o(1).
\end{equation}
\end{lemma}

\begin{proof}
As already mentioned, diagonal elements of $T_\ell(\tfrac 12 J) -
T_\ell(\tfrac 12 J^{(n)})(n)$ vanish starting from $(n+k)^{\rm
th}$. The first $n$ of them are equal to those of $T_\ell(\tfrac
12 J)$, so we are left with proving that the sum of the remaining
$k-1$ is $\tfrac 14(1+(-1)^\ell) + o(1)$.

The $(n+1)^{\rm st}$ through $(n+k-1)^{\rm st}$ diagonal elements
of $T_\ell(\tfrac 12 J)$ differ by $o(1)$ from those of
$T_\ell(\tfrac 12 J_0)$ (since $a_n\to 1$, $b_n\to 0$), and these
are $0$ when $n>k$ \cite[Lemma 3.29]{Zla}. The $(n+1)^{\rm st}$
through $(n+k-1)^{\rm st}$ diagonal elements of $T_\ell(\tfrac 12
J^{(n)})(n)$ differ by $o(1)$ from the $1^{\rm st}$ through
$(k-1)^{\rm st}$ of $T_\ell(\tfrac 12 J_0)$, which sum up to
$-\tfrac 14(1+(-1)^\ell)$ \cite[Lemma 3.29]{Zla}. The proof is
finished.
\end{proof}

With this preparation we can obtain the final form of the sum
rules.

\begin{theorem} \lb{T.5}
Let $a_n\to 1$, $b_n\to 0$, and $w(\tht)=\sum_{\ell=0}^k
c_\ell\cos(\ell\tht)\ge 0$.
\begin{SL}
\item[{\rm{(i)}}] If $f_w\ge 0$ on $[1,1+\eps]\cup[-1-\eps,-1]$
for some $\eps>0$, and either $Z_w(J)<\infty$ or
$\sum_{j=1}^\infty f_w(\beta_j)<\infty$, then $\Tr(P_w(J))$ exists
and
\[
Z_w(J)=\Tr(P_w(J)) + \sum_{j=1}^\infty f_w(\beta_j).
\]
\item[{\rm{(ii)}}] If $\pm f_w\ge 0$ on $[1,1+\eps]$ and $\pm
f_w\le 0$ on $[-1-\eps,-1]$ for some $\eps>0$, and either
$Z_w(J)<\infty$ and $\sum_{\pm E_j\le - 2}f_w(\beta_j)>-\infty$ or
$\sum_{\pm E_j\ge 2}f_w(\beta_j)<\infty$, then $\Tr(P_w(J))$
exists and
\[
Z_w(J) - \sum_{\pm E_j\le -2}f_w(\beta_j) = \Tr(P_w(J)) +
\sum_{\pm E_j\ge 2}f_w(\beta_j).
\]
\item[{\rm{(iii)}}] If $f_w\le 0$ on $[1,1+\eps]\cup[-1-\eps,-1]$
for some $\eps>0$, then $\Tr(P_w(J))$ exists and
\[
Z_w(J)- \sum_{j=1}^\infty f_w(\beta_j) = \Tr(P_w(J)).
\]
\end{SL}
\end{theorem}

{\it Remarks.} 1. The matrix $P_w(J)$ need not be trace-class, but
its trace, given by the sum of its diagonal elements, exists in
(i)--(iii).
\smallskip

2. We use here the convention that $\pm\infty+a=\pm\infty$ for
$a\in\bbR$ and $\infty-\infty$ can be anything. For example, if in
(i) $Z_w(J)<\infty$ and $\sum_{j}f_w(\beta_j)=\infty$, then
$\Tr(P_w(J))$ must be $-\infty$. Notice that in the above sum
rules, $\Tr(P_w(J))$ is the only term that can be $-\infty$.
\smallskip

3. Theorem \ref{T.5}(iii) is just the main result of \cite{NPVY}
in a different guise. It provides a characterization of the
$a_n$'s and $b_n$'s which correspond to matrices with spectral
measures for which a certain Szeg\H o-type integral involving
$\mu'$ and a certain Lieb-Thirring sum are both finite.
\smallskip

4. In the proofs of Theorems \ref{T.1} and \ref{T.1b} we will use
Theorem \ref{T.5}(i) with
\begin{equation} \lb{2.11}
w(\tht)\equiv 3-4\cos 2\tht + \cos 4\tht = 2 (1-\cos 2\tht)^2.
\end{equation}
The previously mentioned results from \cite{KS,Kupin} can be
obtained from Theorem \ref{T.5}(iii) by taking $w(\tht)\equiv
(1-\cos2\tht)^k$ for $k=1,3$, respectively.


\begin{proof}
(i) We take $n\to\infty$ in \eqref{2.3}. Using \eqref{2.5},
\eqref{2.7}, and \eqref{2.10} we obtain
\[
Z_w(J)\ge \limsup_{n\to\infty} \sum_{j=1}^n(P_w(J))_{j,j} +
\sum_{j=1}^\infty f_w(\beta_j).
\]
Similarly, writing \eqref{2.3} for $J_n$ in place of $J$ (with
$(J_n)^{(n)}=J_0$) and taking $n\to\infty$, from \eqref{2.5},
\eqref{2.8}, and \eqref{2.10} we obtain
\[
Z_w(J)\le \liminf_{n\to\infty} \sum_{j=1}^n(P_w(J))_{j,j} +
\sum_{j=1}^\infty f_w(\beta_j).
\]
Here we used the fact that the first $n-k$ diagonal elements of
$P_w(J)$ and $P_w(J_n)$ are the same, whereas the next $k$ differ
by $o(1)$. Unless both $Z_w(J)=\infty$ and $\sum_{j}
f_w(\beta_j)=\infty$, these inequalities can both be satisfied
only if the $\limsup=\liminf$.

The proofs of (ii) and (iii) are analogous.
\end{proof}

\section{Spectral Consequences} \lb{s3}

\begin{proof}[Proof of Theorem \ref{T.1}] First we note that
\begin{equation*} 
T_0(x)=1,\quad T_2(x)=2x^2-1,\quad T_4(x)=8x^4-8x^2+1,
\end{equation*}
and so with $w$ as in \eqref{2.11} we have $\sum_{\ell=0}^k c_\ell
T_\ell(\tfrac x2)=\tfrac 12(4-x^2)^2$. Hence by \eqref{2.1},
$\int_{-2}^2 \ln(\mu'(x)) (4-x^2)^{3/2}\, dx = -\infty$ if and
only if $Z_w(J)=\infty$.


Next, we have
\[
f_w(\beta)\equiv 3\ln(|\beta|)-(\beta^2-\beta^{-2})+\tfrac
1{8}(\beta^4-\beta^{-4}) = \tfrac 8{5}(|\beta|-1)^5 +
O((|\beta|-1)^6).
\]
In particular, $\sum_j(|E_j|-2)^{5/2} = \infty$ if and only if
$\sum_j f_w(\beta_j) = \infty$.

Finally, with $S_{i,j}=\tfrac 78 \delta_{1,i}\delta_{1,j}$, we
have
\[
P_w(J) = S - 3A + 2 T_2(\tfrac 12 J) - \tfrac 1{4}T_4(\tfrac 12 J)
= S - 3A - \tfrac 1{8}(J^4-12J^2+18).
\]
If all $a_j=1$, then for $j\ge 4$ the $j^{\rm th}$ diagonal
element of $P_w(J)$ is
\begin{equation*}
- \tfrac 1{8}   \big[
(b_j^4+6b_j^2+b_{j-1}^2+b_{j+1}^2+2b_j(b_{j+1}+b_{j-1})+6)-12(b_j^2+2)+18
\big].
\end{equation*}
Since $b_j\to 0$, we get that the limit $\Tr(P_w(J))$ exists if
and only if $\sum_{j=1}^\infty [ b_j^4-2(\partial b_j)^2]$ exists,
and
\begin{equation*} 
\Tr(P_w(J)) = - \frac 1{8}\sum_{j=1}^\infty \big[ b_j^4-2(\partial
b_j)^2 \big] + O\big(1+\|b_j\|_\infty\big).
\end{equation*}
For general $d_j\equiv a_j-1$ we get with $O(\|J\|_\infty)=
O(\|a_j\|_\infty+\|b_j|\|_\infty)$,
\begin{align}
-8\Tr(P_w(J)) =   \sum_{j=1}^\infty & \big[ b_j^4-2(\partial
b_j)^2-8(\partial d_j)^2  \notag
\\ +& 8d_j(2d_j^2+d_jd_{j+1}+d_{j+1}^2+b_j^2+b_jb_{j+1}+b_{j+1}^2) \notag
\\ +& 4d_j^2(-d_j^2+d_{j+1}^2+b_j^2+b_jb_{j+1}+b_{j+1}^2) +
O(d_j^5)\big] \notag
\\ +& O \big(\|J\|_\infty \big). \lb{2.12}
\end{align}
If $a_j-1\in\ell^3$, then this is $\sum_{j} r_j + O(\|d_j\|_3^3 +
\|J\|_\infty)$.

(i) From the hypothesis and \eqref{2.12}, we have
$\Tr(P_w(J))=-\infty$ or does not exist. Thus $\sum_j
f_w(\beta_j)=\infty$ by Theorem \ref{T.5}(i) and \eqref{2.2}.

(ii) Now $\Tr(P_w(J))=\infty$ or does not exist, and we use
Theorem \ref{T.5}(i) and $\sum_j f_w(\beta_j)>-\infty$ to get
$Z_w(J)=\infty$.
\end{proof}

By a careful examination of \eqref{2.12}, one can prove the
following variation on Theorem \ref{T.1}, allowing
$a_n-1\notin\ell^3$. We let $a_\pm\equiv\max\{\pm a,0\}$.

\begin{corollary} \lb{C.6}
\begin{SL}
\item[{\rm{(i)}}] If $(a_n-1)_-\in\ell^2$, $\partial a_n,\partial
b_n\in\ell^2$, and either $a_n-1\notin\ell^3$ or
$b_n\notin\ell^4$, then $\sum_j(|E_j|-2)^{5/2} = \infty$.
\item[{\rm{(ii)}}] If $(a_n-1)_+\in\ell^2$, $b_n\in\ell^4$ and
either $a_n-1\notin\ell^3$ or $\partial a_n\notin\ell^2$ or
$\partial b_n\notin\ell^2$, then $\int_{-2}^2 \ln(\mu'(x))
(4-x^2)^{3/2}\, dx = -\infty$.
\end{SL}
\end{corollary}

\begin{proof}
(i) Note that once $d_j$ is small enough, then for $d_j\le 0$ the
sum of the second and third lines in \eqref{2.12} is bounded below
by $-Cd_j^2-\eps(|d_{j+1}|^3+b_j^4+b_{j+1}^4)$ (for any $\eps>0$
and $C=C(\eps)<\infty$), and for $d_j\ge 0$ it is bounded below by
$8d_j^3$. So the whole sum is bounded below by $\sum_j
[(1-2\eps)b_j^4+(8-\eps) d_j^3-q_j]$ for some summable $q_j$,
proving $\Tr(P_w(J))=-\infty$. The result follows as in the proof
of Theorem \ref{T.1}(i).

(ii) One shows that $\Tr(P_w(J))=\infty$ in a similar way as in
(i), this time bounding the sum of the second and third lines of
\eqref{2.12} above by $Cd_j^2+\eps(|d_{j+1}|^3+b_j^4+b_{j+1}^4)$
for $d_j\ge 0$ and by $8d_j^3$ for $d_j\le 0$.
\end{proof}

\begin{proof}[Proof of Theorem \ref{T.1b}]
The hypothesis and \eqref{1.0a} give $\sum_j (|E_j|-2)^{5/2} <
\infty$. Hence, by Theorem \ref{T.5}(i) with $w$ given by
\eqref{2.11}, $\Tr(P_w(J))$ exists. And it is finite if and only
if $Z_w(J)<\infty$. But by the assumptions and \eqref{2.12}, the
former happens precisely when $\partial a_n,\partial
b_n\in\ell^2$. As in Theorem \ref{T.1}, the latter happens if and
only if $\int_{-2}^2 \ln(\mu'(x)) (4-x^2)^{3/2}\, dx > -\infty$.
\end{proof}

Here is an application of Theorem \ref{T.5}(i) to {\it
oscillatory} Jacobi matrices:

\begin{corollary} \lb{C.7}
If $\limsup_n \sum_{j=1}^n (|\partial a_j|^2+|\partial b_j|^2) /
\sum_{j=1}^n (|a_j-1|^3+|b_j|^3) =\infty$, then $\int_{-2}^2
\ln(\mu'(x)) (4-x^2)^{3/2}\, dx = -\infty$.
\end{corollary}

{\it Remark.} This applies, for example, in the case
$a_n=1+\alpha_1\cos(\mu n)/n^{\gamma_1}$ and $b_n=\alpha_2\cos(\mu
n)/n^{\gamma_2}$ when $\mu\notin 2\pi\bbZ$ and $\alpha_j\neq 0$,
$\gamma_j\le\tfrac 12$ for either $j=1$ or $j=2$. In \cite{DHS} it
was proved that in this case, with $\alpha_1=0$, $\int_{-2}^2
\ln(\mu'(x)) (4-x^2)^{-1/2}\, dx = -\infty$.

\begin{proof}
For $w$ as in the proof of Theorem \ref{T.1} we get
$\Tr(P_w(J))=\infty$ (from \eqref{2.12}). Then we continue as in
(ii) of that proof.
\end{proof}

Finally, here are two results illustrating the use of Theorem
\ref{T.5}(ii). The first has been proved in \cite{SZ} for the case
$a_n-1,b_n\in\ell^2$. The second is related to results in
\cite{DHKS,DK}.

\begin{theorem} \lb{T.8}
If $a_n\to 1$, $b_n\to 0$, and $\limsup_n \sum_{j=1}^n
[\ln(a_j)\pm \tfrac 12 b_j] =\infty$, then
\[
\sum_{\pm E_j\ge 2} (|E_j|-2)^{1/2} = \infty.
\]
\end{theorem}

\begin{proof}
Let $w(\tht)\equiv 1\pm\cos\tht$ so that
$f_w(\beta)=2(|\beta|-1)+O((|\beta|-1)^2)$ for $\pm\beta\downarrow
1$ and $f_w(\beta)=-\tfrac 16(|\beta|-1)^3+O((|\beta|-1)^4)$ for
$\pm\beta\uparrow -1$. Then $P_w(J)=-A\mp\tfrac 12 J$, and so its
$j^{\rm th}$ diagonal element is $-[\ln(a_j)\pm \tfrac 12 b_j]$.
Theorem \ref{T.5}(ii) finishes the proof.
\end{proof}

\begin{theorem} \lb{T.9}
If $a_n\to 1$, $b_n\to 0$, and one of $\sum_{n=1}^\infty
[\ln(a_n)\pm \tfrac 12 b_n]$ is larger than $\tfrac 12$ or does
not exist, then $J$ has at least one eigenvalue outside $[-2,2]$.
\end{theorem}

{\it Remark.} The bound $\tfrac 12$ is optimal as can be seen by
taking $a_n\equiv 1$ and $|b_n|\le\delta_{1,n}$. The corresponding
Jacobi matrix has no eigenvalues.

\begin{proof}
Take again $w(\tht)\equiv 1\pm\cos\tht$ in Theorem \ref{T.5}(ii).
By (2.45) in \cite{Zla},
\begin{align*}
Z_w(J) & \ge \int_0^\pi
(1-\cos\tht)\ln(2+2\cos\tht)\,\frac{d\tht}{2\pi}
\\ & = \frac 12 \int_0^{2\pi}
(1-\cos\tht)\ln\big|1+2e^{i\tht}+e^{2i\tht}\big|\,\frac{d\tht}{2\pi}.
\end{align*}
Jensen's formulae for the function $\ln|1+2z+z^2|=\Re(2z+O(z^2))$
show that the last integral equals $-1$.
If $J$ had no eigenvalues, we would have $-\tfrac
12\le Z_w(J)=\Tr(P_w(J))=-\sum_{n=1}^\infty [\ln(a_n)\pm \tfrac 12
b_n]$, a contradiction.
\end{proof}



\end{document}